\documentclass[12pt]{article}
\usepackage{amssymb}
\usepackage{epsfig}
\voffset= - 0.3in
\hoffset= - .6in         

\catcode`\@=11
\def\marginnote#1{}

\newcount\hour
\newcount\minute
\newtoks\amorpm
\hour=\time\divide\hour by60
\minute=\time{\multiply\hour by60 \global\advance\minute by-\hour}
\edef\standardtime{{\ifnum\hour<12 \global\amorpm={am}%
        \else\global\amorpm={pm}\advance\hour by-12 \fi
        \ifnum\hour=0 \hour=12 \fi
        \number\hour:\ifnum\minute<10 0\fi\number\minute\the\amorpm}}
\edef\militarytime{\number\hour:\ifnum\minute<10 0\fi\number\minute}

\def\draftlabel#1{{\@bsphack\if@filesw {\let\thepage\relax
   \xdef\@gtempa{\write\@auxout{\string
      \newlabel{#1}{{\@currentlabel}{\thepage}}}}}\@gtempa
   \if@nobreak \ifvmode\nobreak\fi\fi\fi\@esphack}
        \gdef\@eqnlabel{#1}}
\def\@eqnlabel{}
\def\@vacuum{}
\def\draftmarginnote#1{\marginpar{\raggedright\scriptsize\tt#1}}

\def\draft{\oddsidemargin -.5truein
        \def\@oddfoot{\sl preliminary draft \hfil
        \rm\thepage\hfil\sl\today\quad\militarytime}
        \let\@evenfoot\@oddfoot \overfullrule 3pt
        \let\label=\draftlabel
        \let\marginnote=\draftmarginnote
   \def\@eqnnum{(\theequation)\rlap{\kern\marginparsep\tt\@eqnlabel}%
\global\let\@eqnlabel\@vacuum}  }

\def\appname{Appendix}
\newcounter{app}
\def\theapp{\Alph{app}}
\def\app{\par
   \addvspace{4ex}
   \@afterindentfalse
  \secdef\@app\@dapp}
\def\@app[#1]#2{\ifnum \c@secnumdepth >\m@ne
        \refstepcounter{app}
        \addcontentsline{toc}{app}{\theapp
        \hspace{1em}#1}\else
      \addcontentsline{toc}{app}{ #1}\fi
   {\parindent \z@ \raggedright
    \Large \bf \appname~\theapp .
   \Large  \bf 
    #2}\nobreak
   \vskip 4ex   \noindent
\setcounter{equation}{0}
\def\theequation{\Alph{app}.\arabic{equation}}}
\def\@dapp#1{%
{\parindent \z@ \raggedright  \bf #1}\par\nobreak}
\def\l@app#1#2{\addpenalty{\@secpenalty}%
   \addvspace{1em plus\p@}%
   \begingroup
   \@tempdima 3em
     \parindent \z@ \rightskip \@pnumwidth
     \parfillskip -\@pnumwidth
     { \bf
     \leavevmode
     #1\hfil \hbox to\@pnumwidth{\hss #2}}\par
     \nobreak
   \endgroup}

\makeatletter
\newdimen\normalarrayskip            
\newdimen\minarrayskip               
\normalarrayskip\baselineskip
\minarrayskip\jot
\newif\ifold             \oldtrue            \def\new{\oldfalse}
\def\arraymode{\ifold\relax\else\displaystyle\fi}
\def\eqnumphantom{\phantom{(\theequation)}} 
\def\@arrayskip{\ifold\baselineskip\z@\lineskip\z@
     \else
     \baselineskip\minarrayskip\lineskip1\baselineskip\fi}

\def\@arrayclassz{\ifcase \@lastchclass \@acolampacol \or
\@ampacol \or \or \or \@addamp \or
   \@acolampacol \or \@firstampfalse \@acol \fi
\edef\@preamble{\@preamble
  \ifcase \@chnum
     \hfil$\relax\arraymode\@sharp$\hfil
     \or $\relax\arraymode\@sharp$\hfil
     \or \hfil$\relax\arraymode\@sharp$\fi}}

\def\@array[#1]#2{\setbox\@arstrutbox=\hbox{\vrule
     height\arraystretch \ht\strutbox
     depth\arraystretch \dp\strutbox
width\z@}\@mkpream{#2}\edef\@preamble{\halign \noexpand\@halignto
\bgroup \tabskip\z@ \@arstrut \@preamble \tabskip\z@ \cr}%
\let\@startpbox\@@startpbox \let\@endpbox\@@endpbox
  \if #1t\vtop \else \if#1b\vbox \else \vcenter \fi\fi
  \bgroup \let\par\relax
  \let\@sharp##\let\protect\relax
  \@arrayskip\@preamble}
\def\eqnarray{\stepcounter{equation}%
              \let\@currentlabel=\theequation
              \global\@eqnswtrue
              \global\@eqcnt\z@
              \tabskip\@centering              
              \let\\=\@eqncr
              $$%
            \halign to \displaywidth  \bgroup
             \eqnumphantom \@eqnsel
      \hskip\@centering                               
    $\displaystyle  \tabskip\z@ {##}$%
    &\global\@eqcnt\@ne \hskip 2\arraycolsep
         $ \displaystyle  \arraymode{##}$\hfil
    &\global\@eqcnt\tw@ \hskip 2\arraycolsep
         $\displaystyle\tabskip\z@{##}$\hfil
         \tabskip\@centering
    &{##}\tabskip\z@\cr}
\makeatother

\newfont{\hr}{msbm10}
\newfont{\ams}{msam10}

\font\numbers=cmss12
\font\upright=cmu10 scaled\magstep1
\def\stroke{\vrule height8pt width0.4pt depth-0.1pt}
\def\topfleck{\vrule height8pt width0.5pt depth-5.9pt}
\def\botfleck{\vrule height2pt width0.5pt depth0.1pt}
\def\Zmath{\vcenter{\hbox{\numbers\rlap{\rlap{Z}\kern 0.8pt\topfleck}\kern
2.2pt
                   \rlap Z\kern 6pt\botfleck\kern 1pt}}}
\def\Qmath{\vcenter{\hbox{\upright\rlap{\rlap{Q}\kern
                   3.8pt\stroke}\phantom{Q}}}}
\def\Nmath{\vcenter{\hbox{\upright\rlap{I}\kern 1.7pt N}}}
\def\Cmath{\vcenter{\hbox{\upright\rlap{\rlap{C}\kern
                   3.8pt\stroke}\phantom{C}}}}
\def\Rmath{\vcenter{\hbox{\upright\rlap{I}\kern 1.7pt R}}}
\def\Z{\ifmmode\Zmath\else$\Zmath$\fi}
\def\Q{\ifmmode\Qmath\else$\Qmath$\fi}
\def\N{\ifmmode\Nmath\else$\Nmath$\fi}
\def\C{\ifmmode\Cmath\else$\Cmath$\fi}
\def\R{\ifmmode\Rmath\else$\Rmath$\fi}

\def\d{\partial}

\def\bea{\begin{eqnarray}}
\def\eea{\end{eqnarray}}

\def\beq{\begin{equation}}
\def\eeq{\end{equation}}
\def\ba{\beq\new\begin{array}{c}}
\def\ea{\end{array}\eeq}
\def\be{\ba}
\def\ee{\ea}
\def\F{{\cal F}}

\def\stackreb#1#2{\mathrel{\mathop{#2}\limits_{#1}}}

\def\res{{\rm res}}

\def\half{{\textstyle{1\over2}}}
\def\ha{{1\over 2}}
\def\N2{${\cal N}=2$}
\def\4N{${\cal N}=4$}
\def\1N{${\cal N}=1$}
\def\1N*{${\cal N}=1^*$}

\def\t{{\sf t}}

\def\f{{\sf f}}

\def\beq{\begin{equation}}
\def\eeq{\end{equation}}
\def\ba{\beq\new\begin{array}{c}}
\def\ea{\end{array}\eeq}
\def\be{\ba}
\def\ee{\ea}
\newcommand{\rf}[1]{(\ref{#1})}

\begin{document}

\begin{flushright}
FIAN/TD-03/09\\
ITEP/TH-06/09\\
IHES/P/09/06
\end{flushright}
\vspace{0.3cm}

\renewcommand{\thefootnote}{\fnsymbol{footnote}}
\begin{center}
{\LARGE \bf On two-dimensional quantum gravity and\\
\vspace{0.3cm}
quasiclassical integrable hierarchies~\footnote{
based on talk presented at the conference {\em Liouville field theory and statistical models},
(dedicated to the memory of Alexei B.~Zamolodchikov), Moscow, June 2008}}\\
\vspace{1.5cm}
{\large A.~Marshakov}\\
\vspace{0.5cm}
{\em Theory Department, P.N.Lebedev Physics Institute,\\
Institute of Theoretical and Experimental Physics,\\ Moscow, Russia}\\
\vspace{0.5cm}
{\sf e-mail:\ mars@lpi.ru, mars@itep.ru}\\
\bigskip\bigskip\medskip

\end{center}

\begin{flushright}
{\it to the memory of\\
Alesha Zamolodchikov}\\
\end{flushright}
\vspace{0.3cm}

\begin{center}
\begin{quotation}
\noindent
The main results for the two-dimensional quantum gravity, conjectured
from the matrix model or integrable approach, are presented in the form to be
compared with the world-sheet or Liouville approach. In spherical limit the
integrable side for minimal string theories is completely formulated using
simple manipulations with two polynomials, based on residue formulas from quasiclassical
hierarchies. Explicit computations for particular models are performed
and certain delicate issues of nontrivial relations among them are discussed.
They concern the connections between different theories, obtained as
expansions of basically the same stringy solution to dispersionless KP hierarchy
in different backgrounds, characterized by nonvanishing background values of
different times, being the simplest known example of change of the quantum numbers
of physical observables, when moving to a different point in the moduli space
of the theory.
\end{quotation}
\end{center}
\renewcommand{\thefootnote}{\arabic{footnote}}
\setcounter{section}{0} \setcounter{footnote}{0}
\setcounter{equation}0  \setcounter{page}0
\thispagestyle{empty}

\newpage

\section{Introduction}

The problem of solving two-dimensional quantum gravity exists already more than twenty, or even
more than twenty-five years. By its basic definition one usually takes the Polyakov path integral
\cite{Pol81}, where the integration over the metrics on two-dimensional string world-sheets has been
reduced to study of naively simple, but in fact quite nontrivial two-dimensional conformal
Liouville field theory. The world-sheet approach allowed to determine immediately
only the relatively simple quantities - like scaling dimensions - of the operators of
the two-dimensional quantum gravity \cite{KPZ,DDK}. The computation of their correlators -
even on the world-sheets of simplest spherical topology - appeared to be the problem of much
higher complexity, and was (yet only partially) solved very recently.

Fortunately, the two-dimensional quantum gravity is a {\em renormalizable} theory -
in the most physically important sense of the word, which means that the details of regularization
of the theory at microscopic scale do not affect its macroscopic properties: the ``observable''
scaling dimensions and correlators. In different words, two dimensional quantum gravity
possesses strong {\em universality} property - meaning that quite different methods of the
computation gives rise basically to the same result.

The first sign of this was observed already in the middle of 80-s of the last century.
The idea of summing over the discrete triangulations of world sheets
instead of the integrations
over the metric in continuous theory had demonstrated its efficiency in two-dimensional case,
quite in contrast with the nonrenormalizable gravity of higher dimensions. Moreover, it turned
out that summing over triangulations of the two-dimensional surfaces can be itself reformulated
as summing over the fat graphs of the matrix models \cite{mamo}. The duality between matrix model
(the zero-dimensional gauge theory) and continuous two-dimensional world-sheet gravity is in fact
nothing but the first studied example of the famous nowadays gauge/string duality.

By the matrix model approach two-dimensional quantum gravity was claimed to be ``completely solved''
\cite{ds} in the beginning of 90-s of the previous century. This solution was nicely formulated
\cite{Douglas,FKN} in terms of special stringy solutions to the hierarchies of integrable equations,
being all the
well-known polynomial reductions of the Kadomtsev-Petviashvili (KP) hierarchy. In practice, this
has opened a possibility to compute exactly the correlators in two-dimensional gravity
(in the framework of
``matrix model'' approach) at least in the spherical approximation (when all closed string loops
are suppressed) by methods of dispersionless KP hierarchy, which turn this problem into the problem
of solving algebraic nonlinear equations. Below, following \cite{MM2},
we shall demonstrate how this way leads straightforwardly
to the computation of invariant correlation numbers - the ratios of the correlation functions which
do not depend upon normalizations of particular operators.

However, it is still a great puzzle and, at least partially, an open problem, whether the
matrix model approach leads {\em exactly} to the same results as the original world-sheet approach.
Partially
this is related to the fact, that the world-sheet quantum Liouville theory of \cite{Pol81} is a
rather specific two-dimensional quantum field theory, which is yet to be fully understood. The two-
and three- point functions in Liouville theory were computed in early 90-s \cite{GoLi,DOZZ},
but it turned out, that only after discovery of the higher order equations
of motion by Alesha Zamolodchikov \cite{Zam_ho}, it appeared to be possible to compute the
generic multipoint
correlation functions of the operators of minimal $(p,q)$ models coupled to the
two-dimensional Liouville
gravity, where the integrands on the moduli spaces of world-sheets with punctures
are basically reduced after using the higher order equations to the total derivatives.

These correlation functions could be now compared with the results extracted from the ``matrix model''
approach, or more strictly, from the formulation of minimal string theory in the language of
integrable hierarchies.

\section{dKP for (p,q)-critical points
\label{ss:dkppq}}

According to widely beleived hypothesis, the so called $(p,q)$ critical points of
two-dimensional gravity (or $(p,q)$ minimal string theory) are most effectively described by
the tau-function of $p$-reduced KP hierarchy, satisfying string equation. The logarithm
of this tau-function should be further expanded around
certain background values of the time-variables, with necessary $t_{p+q}\neq 0$.
In particular, it means that the correlators on world-sheets of
spherical topology (the only ones, partially computed by now by means of
two-dimensional conformal field theory \cite{Zam_ho,BeZa}) are governed by quasiclassical
tau-function of dispersionless KP or dKP hierarchy,
which is a very reduced case of generic quasiclassical hierarchy from \cite{KriW}.

For each
$(p,q)$-th minimal theory one should consider a solution of the $p$-reduced dKP hierarchy,
or more strictly, its
expansion in the vicinity of nonvanishing $t_{p+q}={p\over p+q}$
and vanishing other times, perhaps except for cosmological constant $x$, chosen in
a different way for the different theories (the so called conformal backgrounds). If
$q=p+1$ (the unitary series) the cosmological constants $x \sim t_1$ basically coincides
with the main first time of the KP hierarchy, but for ``non-unitary'' backgrounds the quantum
numbers change, and this causes certain nontrivial relations on the space of KP solutions
to be discussed below.

\subsection{Residue formulas}

The geometric formulation of results for minimal string theories in terms of
the quasiclassical hierarchy can be sketched in the following way:

\begin{itemize}
    \item For each $(p,q)$-th point take a pair of polynomials
\be
\label{polspq}
X=\lambda^p+\dots
\\
Y=\lambda^q+\dots
\ee
of degrees $p$ and $q$ respectively. They can be thought of as a dispersionless version of the Lax
and Orlov-Shulman operators of KP theory
\be
\left[{\hat X},{\hat Y}\right]=\hbar
\\
{\hat X} = \d^p + \ldots,
\ \ \ \
{\hat Y} = \d^q + \ldots
\ee
or as a pair of (here already integrated) Krichever differentials with the fixed
periods on a complex curve (for dKP - a rational curve with global uniformizing
parameter $\lambda$). It is also convenient to combine these polynomials into
a generating differential
\be
\label{dSpq}
dS = YdX
\ee
whose periods and singularities define the variables of the quasiclassical hierarchy.
Since on rational curve ($\lambda$-plane or Riemann sphere with the marked point $P_0$, where
$\lambda=\infty$) all periods of \rf{dSpq} vanish, the time variables are related with the
residues or singular part of expansion of the differential $dS$ at the point $P_0$.
\item
The variables of dispersionless KP hierarchy are therefore introduced by residue formulas
\cite{KriW,LGGKM,TakTak}
\be
\label{tP}
t_k = {1\over k}\ \res_{P_0} \xi^{-k}dS,\ \ \ k>0
\\
{\d\F\over \d t_k} =  \res_{P_0} \xi^{k}dS,\ \ \ k>0
\ee
where
\be
\label{locop}
\xi = X^{1\over p} = \lambda\left(1+
\dots + {X_0\over\lambda^p}\right)^{1\over p}
\ee
is the distinguished inverse local co-ordinate at the
point $P_0$, where $\lambda(P_0)=\infty$ and $\xi (P_0) = \infty$.
From \rf{tP} it also follows for the second derivatives
\be
\label{sysi}
{\d^2\F\over \d t_n\d t_k} = \res_{P_0} (\xi^k dH_n)
\ee
while the third derivatives are given by the formula
\be
\label{residue}
{\d^3 \F\over \d t_k\d t_l\d t_n} =
\res_{dX=0}\left(dH_kdH_ldH_n\over dX dY\right)
\ee
In \rf{sysi} and \rf{residue} the set of one-forms
\be
\label{smer}
dH_k={\d dS\over \d t_k} , \ \ \ k\geq 1
\ee
(derivatives are taken at fixed $X$)
corresponds to dispersionless limit of KP flows and can be integrated up to
polynomial expressions
\be
\label{Hpol}
H_k = X(\lambda)^{k/p}_+
\ee
in uniformizing co-ordinate
$\lambda = H_1$.
\end{itemize}
Note also, that the tau-functions of $(p,q)$ and $(q,p)$ theories do not
coincide, but are related by the Legendre or Fourier transform \cite{KhMa},
exchanging the polynomials \rf{polspq} by each other $X \leftrightarrow Y$.

\subsection{Solution to dKP}

The fact, that one-forms \rf{smer} can be integrated up to the polynomials \rf{Hpol}
leads to explicit expression for the integrated generating differential \rf{dSpq}, or
\be
\label{Spkp}
S = \sum_{k=1}^{p+q} t_k H_k
= \sum_{k=1}^{p+q} t_k X^{k/p}(\lambda)_+,\ \ \ \ k\ {\rm mod}\ p
\ee
depending already upon
the coefficients of the polynomial $X(\lambda)$ only. In different words, formula \rf{Spkp}
means, that the first part of the equations \rf{tP} has been already effectively
resolved for the coefficients of $Y(\lambda)$.
The dependence of coefficients of $X(\lambda)=\lambda^p + \sum_{k=0}^{p-2}X_k\lambda^k$
over the KP times \rf{tP} is determined in the most easy way from $\left. dS\right|_{dX=0} = 0$,
which is now a system
of $p-1$ ``hodograph'' equations ${dS\over d\lambda}=0$ imposed at $p-1$ roots of $X'(\lambda)=0$.

A simple observation, that any hamiltonian \rf{Hpol} is a polynomial in terms of the
variable $\lambda=H_1$
leads to dispersionless Hirota equations, which express any second derivative
${\d^2\F\over \d t_k\d t_n}$ with arbitrary $k$ and $l$ in terms of the second derivatives
${\d^2\F\over \d t_k\d t_1}$ where one of the indices is fixed and corresponds to the first
time. From formulas \rf{tP} one finds that
\be
\label{yx}
dS \stackreb{\xi\to\infty}{=}  \sum\left( kt_k \xi^{k-1}d\xi + {\d \F\over\d t_k}{d\xi\over
\xi^{k+1}}\right)
\ee
which is just an expansion in local co-ordinate at the marked point $P_0$.
Taking the time derivatives (cf. with \rf{smer}) gives the set
\be
\label{omkp}
H_k = {\d S\over\d t_k} = \xi^k - \sum_j
{\d^2 \F\over\d t_k\d t_j}{1\over j\xi^j} = \xi^k(\lambda)_+,
\ \ \ \ k>0
\ee
which forms a basis of meromorphic functions with poles at the point
$P_0$, or just a particular polynomial basis, explicitly fixed by last equation.
The set of the powers $\lambda^k$ has the same singularities as the set
of functions (\ref{omkp}), i.e. these two are related by simple
linear transformation, e.g.
\be
\label{pomeg}
H_1 = \lambda, \ \ \ \
H_2 = \lambda^2 + 2{\d^2 \F\over\d t_1^2},
\\
H_3 = \lambda^3 + 3{\d^2 \F\over\d t_1^2}\lambda +
{3\over 2}{\d^2 \F\over\d t_1\d t_2}, \ \ \ \dots
\ee
These equalities follow from the comparison of the singular at $P_0$
part of their expansions in $\xi$, following from (\ref{omkp}). Comparing
the negative
"tails" of the expansion in $\xi$ of both sides of eq.~(\ref{pomeg}) expresses
the derivatives
${\d^2 \F\over\d t_k\d t_l}$ (of $H_k$ in the l.h.s.) in terms of only
those with $k=1$ (of $\lambda=H_1$ in the r.h.s.). These relations are called
the dispersionless KP, or the dKP Hirota equations, e.g.
\be
\label{252d}
{\d^2\F\over\d t_1\d t_3} = {3\over 8}X_0^2 =
{3\over 2}\left({\d^2\F\over\d t_1^2}\right)^2
\\
{\d^2\F\over\d t_3\d t_3} = {3\over 8}X_0^3 =
3\left({\d^2\F\over\d t_1^2}\right)^3
\ee
We have listed here those, which will be of some interest for two-dimensional
quantum gravity.

\subsection{Scaling}

Under the scaling $X\to\Lambda^p X$, $Y\to\Lambda^q Y$,
(induced by $\lambda\to\Lambda\lambda$ and therefore
$\xi\to\Lambda\xi$), the times \rf{tP}
transform as $t_k \to\Lambda^{p+q-k}t_k$. Then from the second formula
of \rf{tP} it follows that the function $\F$ scales as $\F\to\Lambda^{2(p+q)}\F$,
or, for example, as
\be
\label{scaF}
\F \propto t_1^{\ 2{p+q\over p+q-1}}\ f(\tau_k)
\ee
where $f$ is supposed to be a scale-invariant function of corresponding
dimensionless ratios of the times $\tau_k=t_k/t_1^{p+q-m\over p+q-1}$ \rf{tP}.
In the simplest $(p,q)=(2,2K-1)$ case of dispersionless KdV one also
expects a natural scaling of the form
\be
\label{scakdv}
\F \propto \left(t_{2K-3}\right)^{K+\half}\ \f({\sf t}_l)
\ee
with ${\sf t}_l = t_{2l-1}/(t_{2K-3})^{(K-l+1)/2}$, where the role of cosmological
constant is played by the time $t_{2K-3}\propto \Lambda^4$.

\subsection{KdV series}

More explicit formulas can be written for the ``KdV-series'' $(p,q)=(2,2K-1)$,
corresponding to the $p=2$ KdV reduction of the KP hierarchy. Now
\be
\label{polkdv}
X=\lambda^2+2u,\ \ \ \xi=\sqrt{X}=\sqrt{\lambda^2+2u}
\\
Y=\lambda^{2K-1}+\sum_{k=1}^{K-1} y_k\lambda^{2k-1}
\ee
and the explicit formula \rf{Spkp} reads
\be
\label{Skdv}
S = \sum_{k=1}^{K+1} t_{2k-1}X^{k-1/2}(\lambda)_+
\ee
Dependence on $u$ upon the flat times is
determined by a {\em single} equation
\be
\label{Krieq}
\left. dS\right|_{dX=0} = 0
\ee
since $dX=2\lambda d\lambda$ has the only zero at $\lambda=0$,
or vanishing of the polynomial
\be
\label{godeq}
P(u) \equiv
\half \left.{dS\over d\lambda}\right|_{\lambda=0} = \sum_{k=0}^K {(2k+1)!!\over k!}t_{2k+1}u^k = 0
\ee
Integrating square of the polynomial \rf{godeq}
\be
\label{F2P}
\F = \half\int_0^u P^2(v)dv = \half\sum_{k,l=0}^K t_{2k+1}t_{2l+1}
{(2k+1)!!(2l+1)!!\over k!l!(k+l+1)}u^{k+l+1}
\ee
one gets the string free energy - the logarithm of quasiclassical tau-function,
due to the formula
\be
\label{tplus0}
\F = \half\sum_{k,l} t_kt_l\ \res_{P_0} (\xi^k dH_l)
\ee
expressing free energy \cite{LGGKM} in terms of its second derivatives, and
since the coefficient in the r.h.s. of \rf{F2P} exactly coincides with the second derivative
\rf{sysi}
\be
\label{sykdv}
\res_{\lambda=\infty} (\xi^{2k+1} dH_{2l+1}) =
\\
= \sum_{n\geq 0}\sum_{m=0}^l
{(2u)^{n+m}\over n!m!}{\Gamma(k+3/2)\Gamma(l+3/2)(2(l-m)+1)\over\Gamma(k+3/2-n)\Gamma(l+3/2-m)}
\res_{\lambda=\infty}\left(d\lambda \lambda^{2(k+l-n-m)+1}\right) =
\\
= (2u)^{k+l+1}\Gamma(k+3/2)\Gamma(l+3/2){2\over\pi}\sum_{m=0}^l{(-)^{l-m}\over m!(k+l+1-m)!} =
\\
= {(2k+1)!!(2l+1)!!\over k!l!(k+l+1)}u^{k+l+1}
\ee
where the last equality holds, in particular, due to binomial identity
$\sum_{m=0}^l(-)^m(^s_n)=(-)^l(^{s-1}_{\ l})$.

\section{Examples: particular (p,q) models}

\subsection{Pure gravity: the explicit partition function
\label{ss:23}}

In this case $(p,q)=(2,3)$, one has only two nontrivial parameters $t_1$ and $t_3$, and
the partition function can be calculated explicitly. The times \rf{tP} are expressed by
\be
\label{t23}
t_5={2\over 5}, \ \ \ t_3={2\over 3}Y_1-X_0, \ \ \ t_1={3\over 4}X_0^2-X_0Y_1
\\
t_4=\ha Y_2,\ \ \ t_2=Y_0-Y_2X_0
\ee
in terms of the coefficients of the
polynomials
\be
\label{xy23}
X=\lambda^2+X_0, \ \ \ \ Y=\lambda^3+ Y_2\lambda^2 + Y_1\lambda + Y_0
\ee
The odd times $t_1$, $t_3$ and $t_5$ do not depend upon the even coefficients
$Y_0$ and $Y_2$ of the second polynomial in \rf{xy23}, and in what follows we choose
$Y_2=Y_0=0$, ensuring $t_2=t_4=0$.
Relations \rf{t23} can be then easily solved for the latter coefficients of
\be
\label{xy23red}
X=\lambda^2+X_0, \ \ \ \ Y=\lambda^3+ Y_1\lambda
\ee
giving rise to
\be
\label{coef23}
X_0={1\over 3}\sqrt{9t_3^2-12t_1}-t_3,
\ \ \ \
Y_1={1\over 2}\sqrt{9t_3^2-12t_1}
\ee
The second half of residues \rf{tP} gives
\be
\label{f123}
{\d\F\over\d t_1} = {1\over 8}X_0^3-{1\over 4}Y_1X_0^2
\\
{\d\F\over\d t_3} = -{1\over 8}Y_1X_0^3+{3\over 64}X_0^4
\ee
This results in the following explicit formula for the quasiclassical tau-function
\be
\label{Fpugr}
\F = {1\over 3240}\left(9t_3^2-12t_1\right)^{5/2}+
{1\over 4}t_3^3t_1-{1\over 4}t_3t_1^2-{3\over 40}t_3^5
\ee
At $t_3\to\infty$ (expansion at $t_1\to 0$) formula \rf{Fpugr} gives
\be
\label{konlim}
\F\ \stackreb{t_3\to\infty}{=}\ - {t_1^3\over 18t_3}\left(1 +
O\left({t_1\over t_3^2}\right)\right)
\ee
which is the partition function of the Kontsevich model \cite{Kontsevich,GKM}
(also identified with the $(2,1)$-point or
topological gravity). At $t_1\to\infty$ tau
function \rf{Fpugr} scales as $\F \propto t_1^{5/2}$ or partition function
of pure two-dimensional gravity: expansion at $t_1\to\infty$ gives
\be
\F = (-3t_1)^{5/2}\left({4\over 405}-{1\over 54}{t_3^2\over t_1}+{1\over 96}
{t_3^4\over t_1^2}+O\left({t_3^6\over t_1^3}\right)\right) + \ldots
\ee
modulo analytic terms.

Formula \rf{Fpugr} is the only example of {\em exact} computation. For the rest one needs to
solve perturbatively the nonlinear string equation. It contains the
polynomial part, which gives contribution only to a finite number of correlation functions.
Usually such ``non-universal'' part is neglected, when comparing the result of the computation
with the world-sheet Liouville theory. It also vanishes at $t_3=0$ or at vanishing of the time,
corresponding to so called ``boundary operator'' (see e.g. \cite{MSS}), the $t_{2K-1}$ variable
in the $(2,2K-1)$ KdV series, which we shall usually neglect in what follows. However, these terms
are essential, when taking the limit \rf{konlim} to the topological Kontevich model, and it means
that they come from the contact terms of topological origin.

\subsection{The gravitational Yang-Lee model: (p,q)=(2,5)
\label{ss:25}}

The calculation of times according to \rf{tP} gives
\be
\label{txy25}
t_1 = -{5\over 8}X_0^3+{3\over 4}Y_3X_0^2-Y_1X_0
\\
t_3= {5\over 4}X_0^2-Y_3X_0+{2\over 3}Y_1
\\
t_5 = {2\over 5}Y_3-X_0
\\
t_7 = {2\over 7}
\ee
for the polynomials
\be
\label{pol25}
X= \lambda^2+X_0
\\
Y= \lambda^5+Y_3\lambda^3+Y_1\lambda
\ee
These equations are easily solved for
\be
\label{yj}
Y_1={3\over 2}\left(t_3+{5\over 4}X_0^2+{5\over 2}t_5X_0\right)
\\
Y_3 = {5\over 2}\left(X_0+t_5\right)
\ee
ending up with the only nonlinear string equation for $X_0$
\be
\label{gpse}
t_1 = -{5\over 8}X_0^3-{3\over 2}t_3X_0
\ee
The one-point functions \rf{tP} are given by
\be
\label{1p25}
{\d\F\over \d t_1} = -{15\over 64}X_0^4 - {3\over 8}t_3X_0^2
\\
{\d\F\over \d t_3} = -{9\over 64}X_0^5 - {3\over 16}t_3X_0^3
\ee
while the two-point functions are
\be
\label{252dd}
{\d^2\F\over\d t_1^2} = {X_0\over 2},
\ \ \ \
{\d^2\F\over\d t_1\d t_3} = {3\over 8}X_0^2,
\ \ \ \
{\d^2\F\over\d t_3\d t_3} = {3\over 8}X_0^3
\ee
The last expressions can be obtained by differentiation \rf{1p25} upon following from
\rf{gpse} explicit formulas for ${\d X_0\over\d t_1}$ and ${\d X_0\over\d t_3}$, or
they follow directly from the Hirota equations \rf{252d}.

To compare the predictions of the ``integrable'' approach for correlators
in two-dimensional gravity with the calculations in world-sheet theory, one needs
first to make certain correspondences in the space of coupling constants.
The simplest one comes from the scaling properties \rf{scaF}, \rf{scakdv}.
In the Yang-Lee theory the role of the cosmological constant is played by the KdV
time $t_3$, and from the scaling properties of the ``fixed area'' partition
function $F_A(t_1)=A^{-7/2}{\sf z}\left(t_1A^{3/2}\right)$ (cf. with \cite{Zamol}) one
gets for the Laplace transformed
$\F (t_1,t_3) = \int_0^\infty {dA\over A}e^{-t_3A}F_A(t_1)$ , or
\be
\label{25sca1}
\F = t_3^{7/2}\ {\sf f}\left({t_1\over t_3^{3/2}}\right)
\equiv t_3^{7/2}\ {\sf f}(\t)
\\
{\d\F\over \d t_1} = t_3^2{\sf f}', \ \ \ \ \
{\d^2\F\over \d t_1^2} = t_3^{1/2}{\sf f}'',\ \ \ldots
\ee
and the string equation turns into
\be
\label{gpsca}
{\sf t} +5(\f'')^3 + 3\f'' = 0
\ee
to be solved for the coefficients
$\f_n \equiv \left.\f^{(n)}\right|_{\t=0}$
in the expansion of
\be
\label{exp25norm}
\F = t_3^{7/2}\f_0 + t_1t_3^2\f_1 + {t_1^2t_3^{1/2}\over 2}\f_2
+ {t_1^3\over 6t_3}\f_3 + \ldots
\ee
which gives rise to rational expressions
\be
\label{fsca}
{\sf f}_3 = -{1\over 3(1+5{\sf f}_2^2)},
\ \ \
{\sf f}_4 = -{10{\sf f}_2\over 9(1+5{\sf f}_2^2)^3},
\ \ \
{\sf f}_5 = {10(1-25{\sf f}_2^2)\over 27(1+5{\sf f}_2^2)^5}
\\
\f_6 = {1000\f_2(1-10\f_2^2)\over 81(1+5\f_2^2)^7},
\ \ \
\f_7 = - {1000(1-95\f_2^2+550\f_2^4)\over 243(1+5\f_2^2)^9}
\\
\f_8 = -{70000\f_2(2-70\f_2^2+275\f_2^4)\over 2187(1+5\f_2^2)^{11}},
\ \ \
\ldots
\ee
in terms of the two-point function $\f_2$, which itself can be found as a
nonvanishing solution to the ``reduced'' string equation
\be
3{\sf f}_2 + 5\f_2^3 = 0
\ee
The ``total normalization'' ${\sf f}_0$ and the ``one-point function'' ${\sf f}_1$, which
does not have a universal sense, since it is coupled to an analytic term in the expansion
\rf{exp25norm}, in principle are
determined by residue formula for ${\d\F/\d t_3}$, or
\be
7\f - 3\t\f'+9(\f'')^5 + 3(\f'')^3=0
\ee
giving rise to
\be
\f_0 = -{9\over 7}\f_2^5-{3\over 7}\f_2^3 = -{3\over 7}\f_2^3\left(1+3\f_2^2\right)
\\
\f_1 = -{9\over 4}\f_2^3\f_3-{45\over 4}\f_2^4\f_3 = {3\over 4}\f_2^2
\ee
This results in the rational ``invariant ratios'', e.g.
\be
\label{ratsca}
{{\sf f}_4{\sf f}_2\over {\sf f}_3^2} = -3,\ \ \
{{\sf f}_4{\sf f}_3\over {\sf f}_2{\sf f}_5} = -{1\over 8}, \ \ \
{\f_2\f_4\over\f_0\f_6} = 1,\ \ \
{\f_4^2\over\f_0\f_8} = -{6\over 143},
\\
{\f_2^2\over\f_0\f_4} = {\f_2\f_6\over\f_4^2} = -35
\ee
to be possibly compared with the computations in the world-sheet theory.

\subsection{Mixing in $(2,7)$ model
\label{ss:27}}

The $(p,q)=(2,7)$ model naively is not much different from the
Yang-Lee case of $(2,5)$ theory considered in sect.~\ref{ss:25}.
The polynomials \rf{polspq} are
\be
\label{pol27}
X= \lambda^2+X_0
\\
Y= \lambda^7 + {7X_0\over 2}\lambda^5+Y_3\lambda^3+Y_1\lambda
\ee
and the calculation of flat times \rf{tP} gives
\be
\label{txy27}
t_1 = {3\over 4}Y_3X_0^2-{105\over 64}X_0^4-Y_1X_0
\\
t_3= {35\over 12}X_0^3-Y_3X_0+{2\over 3}Y_1
\\
t_5 = -{7\over 4}X_0^2+{2\over 5}Y_3
\\
t_7 = 0
\\
t_9 = {2\over 9}
\ee
Again we see, that \rf{txy27} can be easily solved w.r.t. $Y_j$, but
the only coefficient $X_0$ now satisfies
\be
\label{se27}
t_1 = -{35\over 64}X_0^4-{15\over 8}t_5X_0^2-{3\over 2}t_3X_0
\ee
where we put $t_7=0$ for the coefficient at the "boundary" operator \cite{MSS}.

The one-point functions \rf{tP} are given for the $(2,7)$ model by
\be
\label{1p27}
{\d\F\over \d t_1} =
-{7\over 32}X_0^5-{1\over 4}Y_1X_0^2+{1\over 8}Y_3X_0^3
= -{7\over 32}X_0^5 - {5\over 8}X_0^3t_5-{3\over 8}X_0^2t_3
\\
{\d\F\over \d t_3} =
-{1\over 8}Y_1X_0^3-{35\over 512}X_0^6+{3\over 64}Y_3X_0^4
= -{35\over 256}X_0^6-{45\over 128}X_0^4t_5-{3\over 16}X_0^3t_3
\\
{\d\F\over \d t_5} =
-{15\over 512}X_0^7-{5\over 64}Y_1X_0^4+{3\over 128}Y_3X_0^5 =
-{25\over 256}X_0^7-{15\over 64}X_0^5t_5-{15\over 128}X_0^4t_3
\ee
In the r.h.s.'s of \rf{1p27} we already substituted the expressions for
$Y_j$ in terms of times \rf{txy27}, and the rest is to solve \rf{se27} by
expanding in $t_3$ and $t_5$ and substitute result into \rf{1p27}.

The scaling anzatz \rf{scakdv}, \rf{25sca1} now reads
\be
\label{27sca1}
\F = t_5^{9/2}\ {\sf f}\left({t_1\over t_5^2},{t_3\over t_5^{3/2}}\right)
\\
{\d\F\over \d t_1} = t_5^{5/2}{\sf f}^{(1)}, \ \ \ \ \
{\d^2\F\over \d t_1^2} = {X_0\over 2} = t_5^{1/2}{\sf f}^{(11)},\ \ \ldots
\ee
where we have introduced shorten notation for the derivatives over the first argument
of $\f(\t_1,\t_2)$,
and string equation \rf{se27} turns into
\be
\label{27sca}
{\sf t}_1 +{35\over 4}{\sf u}^4 + {15\over 2}{\sf u}^2 + 3\t_2{\sf u} = 0
\ee
for ${\sf u} = {\sf f}^{(11)}$.

Expansion should be considered in the vicinity of the point $t_1={25\over 28}t_5^2$, where
the one-point function in the first equation of \rf{1p27} vanishes on string equation \rf{se27}
at $t_3=0$. It means, in particular, that the function ${\sf f}$ should be expanded around the
non-vanishing background value ${\sf t}_1={25\over 28}$ of its first argument.

\subsection{Ising model (p,q)=(3,4)}

The residue formulas for the polynomials
\be
\label{ispol}
X = \lambda^3 + X_1\lambda + X_0
\\
Y = \lambda^4 + Y_2\lambda^2 + Y_1\lambda + Y_0
\ee
give rise to
\be
\label{coef34}
Y_2 = {4\over 3}X_1 + {5\over 3}t_5
\\
Y_0= {2\over 9}X_1^2 + {10\over 9}X_1t_5
\\
Y_1= {4\over 3}X_0 
\ee
(where the last equation is true upon $t_4=0$), while $X_0$ and $X_1$ satisfy
\be
\label{seis}
t_1=-{2\over 3}X_0^2+{4\over 27}X_1^3 +{5\over 9}t_5X_1^2 
\\
t_2=-{2\over 3}X_0X_1-{5\over 3}t_5X_0 
\ee
Differentiating equations \rf{seis} one can find explicitly expressions for the first
derivatives
\be
\label{Xisder}
{\d X_1\over \d t_j} = {Q_1^{(j)}\over R},\ \ \ {\d X_0\over \d t_j} = {Q_0^{(j)}\over R},
\ \ \ j=1,2,5
\ee
with $R=4X_1^3+12X_0^2+20t_5 X_1^2+25t_5^2X_1$ and
\be
\label{Qpol}
Q_1^{(1)}={9\over 2}(2X_1+5t_5) ,\ \ \ Q_0^{(1)}=-9X_0
\\
Q_1^{(2)}=-18X_0 ,\ \ \ Q_0^{(2)}=-3X_1(2X_1+5t_5)
\\
Q_1^{(5)}=-{5\over 2}(2X_1^3+5t_5X_1^2+12X_0^2) ,\ \ \ Q_0^{(5)}=-5X_1X_0(X_1+5t_5)
\ee
Solving the second equation of \rf{seis} for $X_0$ and substituting the result into the first one,
turns it into the Boulatov-Kazakov equation for $X_1$ \cite{BuKa}
\be
\label{kbeq}
t_1=-{6t_2^2\over (2X_1+5t_5)^2}+{4\over 27}X_1^3 +{5\over 9}t_5X_1^2
\ee
(contains information about all singularities of $\F$
for arbitrary magnetic field $t_2$ and fermion mass $t_5$).

It is interesting to compare the Boulatov-Kazakov equation with what gives here formula
\rf{Krieq}. The branch points are given by $dX=0$ for the first polynomial from \rf{ispol}, or
$\lambda_\pm = \pm \sqrt{-{X_1\over 3}}$, so that vanishing of the derivative of function
\be
\label{Sis}
S = t_1X(\lambda)^{1/3}_+ + t_2X(\lambda)^{2/3}_+ + t_5X(\lambda)^{5/3}_+
+ {3\over 7}X(\lambda)^{7/3}_+
\ee
at $\lambda_\pm$ or $\left. S'(\lambda)\right|_{\lambda_+}=\left. S'(\lambda)\right|_{\lambda_-}$
gives rise to the last equation of \rf{seis}, to be easily solved for $X_0$. Substituting the
result into $\left. S'(\lambda)\right|_{\lambda_+}+\left. S'(\lambda)\right|_{\lambda_-}=0$
reproduces immediately the string equation \rf{kbeq}.

The one-point functions
\be
\label{ising1}
{\d\F\over \d t_1} = {1\over 27}X_1^4
+{10\over 81}t_5X_1^3 -{4\over 9}X_1X_0^2 - {5\over 9}t_5X_0^2
\\
{\d\F\over \d t_2} = {4\over 27}X_1^3X_0+{10\over 27}t_5X_1^2X_0-{8\over 27}X_0^3
\\
{\d\F\over \d t_5} = {40\over 243}X_1^3X_0^2-{10\over 2187}X_1^6
+{25\over 81}t_5X_1^2X_0^2-{10\over 729}t_5X_1^5-{5\over 27}X_0^4
\ee
give rise to
\be
\label{uis}
{\d^2\F\over \d t_1^2} = {X_1\over 3}, \ \ \ {\d^2\F\over \d t_1\d t_2} = {2X_0\over 3},
\\
{\d^2\F\over \d t_1\d t_5} = {5\over 9}X_0^2-{5\over 81}X_1^3,
\ \ \ {\d^2\F\over \d t_2\d t_5} = -{10\over 27}X_1^2X_0,
\\
{\d^2\F\over \d t_2^2} = -{2\over 9}X_1^2,\ \ \
{\d^2\F\over \d t_5^2} = - {50\over 81}X_1^2X_0^2 + {5\over 243}X_1^5,
\ \ \ \ldots
\ee
At $t_2=0$ one gets for the one-point functions \rf{ising1}
\be
\label{ising1}
\left.{\d\F\over \d t_1}\right|_{t_2=0} = {1\over 27}X_1^4
+{10\over 81}t_5X_1^3,
\ \ \ \
\left.{\d\F\over \d t_2}\right|_{t_2=0} = 0
\\
\left.{\d\F\over \d t_5}\right|_{t_2=0} = -{10\over 729}X_1^5\left(t_5 + {X_1\over 3}\right)
\ee
Note also, that for $t_2=0$ the second equation of \rf{seis} has the only reasonable
solution $X_0=0$, while the first one turns into
\be
\label{isred}
t_1={4\over 27}X_1^3 +{5\over 9}t_5X_1^2
\ee
which almost coincides with the perturbation of the Yang-Lee $(2,5)$ model by quadratic term
(cf. with \rf{gpse} and note that potential ${X_1\over 3}$ from \rf{isred} is an analog of
${X_0\over 2}$ from \rf{gpse}, see \rf{252dd} and \rf{uis}). More strictly,
the quadratic term can be removed by redefinition
\be
\label{resisi}
{\hat t}_1 = {4\over 27}{\hat X}_1^3 - {25\over 36}t_5^2{\hat X}_1
\\
{\hat t}_1=t_1-{125\over 216}t_5^3,\ \ \ {\hat X}_1 = X_1+{5\over 4}t_5
\ee
This redefinition exactly fits \cite{MSS} the
vanishing of the energy 3-point function in the
Ising model. Indeed, if one identifies ${\hat t}_1$ with the cosmological constant of the
world-sheet theory, the energy 3-point functions
\be
\label{e3isi}
\left.{\d^3\over \d t_5^3}\F ({\hat t}_1+Ct_5^3,t_2=0,t_5)\right|_{t_5=0} =
\left.\left( 6C{\d\F\over\d t_1}+{\d^3\F\over\d t_5^3}\right)\right|_{t_{2,5}=0}
\ee
vanishes exactly at $C={125\over 216}$. To calculate the r.h.s. of \rf{e3isi} one can
use the first equation from \rf{ising1} and differentiate the last formula from \rf{uis}
using \rf{Xisder}, which is quite easy since $X_0=0$ at $t_2=0$. An alternative and more
fundamental way is to use directly the residue formula \rf{residue} for the third derivatives,
which gives here
\be
\label{res555}
{\d^3\F\over\d t_5^3} = \res_{dX=0}\left(dH_5^3\over dX dY\right) =
\sum_{\lambda=\lambda_\pm}{H_5'(\lambda)^3\over 6\lambda Y'(\lambda)}\
\stackreb{t_2=t_5=0}{=}\ - {125\over 972}X_1^4
\ee
It is interesting to point out, that
under reparameterization \rf{resisi} in the space of couplings
\be
\label{renoc}
X_1 = {\hat X}_1 - {5\over 4}t_5
\\
t_1 = {\hat t}_1 + {125\over 216}t_5^3
\ee
the reduced string equation \rf{isred} acquires the form of (analytically continued)
string equation \rf{gpse} for the Yang-Lee model, with $t_3^{\rm YL}\sim t_5^2$ of
the Yang-Lee model being substituted by the {\em square} of the $t_5=t_5^{\rm Ising}$
of the (reduced) Ising model. However, one should use the scaling anzatz \rf{scaF} rather than
\rf{scakdv}, has been used for the $(2,5)$ theory in \rf{25sca1}, for the function
$F({\hat t}_1,t_5)=\left.\F({\hat t}_1 + {125\over 216}t_5^3,t_5)\right|_{t_2=0}$.
Since
\be
\label{FhX1}
{\d^2 F\over \d{\hat t}_1^2} = {\hat X}_1({\hat t}_1,t_5^2) - {5\over 4}t_5
\ee
as follows from string equation \rf{resisi} for ${\hat X}_1$ and the couplings are dimensional,
the gravitational Ising free energy $F = {\hat F}({\hat t}_1,t_5^2) - {5\over 8}t_5{\hat t}_1^2$
is an even function of $t_5$, apart of an analytic cubic term, and
its expansion
gives all the $\langle\epsilon^{2n}\rangle$
correlators of the gravitationally dressed $(3,4)$ Ising model. We shall comment more
about the relation of these two models in the next section.

\section{Ising versus Yang-Lee}

Both gravitational Ising model and $(2,5)$ Yang-Lee minimal theory arise as two different
critical points in a system of Ising spins on a random lattice. Moreover, since both theories
have $p+q=7$, they have identical scaling in the first KP variable \rf{scaF}, which originally
has cause a confusion, when distinguishing these two minimal string theories. In particular,
that origins from the fact that string equations of those two models can be obtained from each
other by simple reparameterization in the space of couplings, as we have noticed already in the
previous section.

However, the physical sense of parameters, arising in these two equations is totally different.
One can say, that the same KP time variable has different ``quantum numbers'', when one takes
a solution, corresponding to a different critical point. For example, the role of cosmological
constant, coupled to a unity operator on world-sheet is played by $t_1$ in $(3,4)$ Ising, but
by $t_3$ in the $(2,5)$ Yang-Lee theory. Below we shall try to present more details about this
relation and describe it as much as possible from the point of view of (dispersionless) KP theory.

\subsection{Kostov equation}

This is a name, given by Alesha Zamolodchikov to a ``phenomenological''
transcendental equation, satisfied by the second derivative of free energy over the
cosmological constant $u\sim{\d^2\F\over\d x^2}$ of the form
\be
\label{Koeq}
u^\nu + tu^{\nu-1}=x
\ee
where $\nu =\nu (p,q) = {p\over q-p}$. For the cases of interest one gets integer
$\nu(2,3)=2$ for pure gravity and $\nu(3,4)=3$ for Ising (both are unitary with $q=p+1$), but
$\nu(2,5)={2\over 3}$.

Hence, for the Ising model the Kostov equation reads
\be
\label{kois}
u_I^3 +t_Iu_I^2=x_I
\ee
and coincides (after renormalization $x_I\sim t_1$, $t_I\sim t_5$ and
$u_I\sim X_1 = 3{\d^2\F\over\d t_1^2}$, below in this section
we shall use different normalization from conventional in KP theory, to get rid of ugly numerical
constants) with the Boulatov-Kazakov equation \rf{isred} when $t_2=0$, i.e. for vanishing magnetic
field.

For the Yang-Lee model equation \rf{Koeq} $u_{YL}^{2/3}+t_{YL}u_{YL}^{-1/3}=x_{YL}$ after the substitution
$u_{YL}=v_{YL}^3$ turns into
\be
\label{koyl}
v_{YL}^3 - x_{YL}v_{YL}=-t_{YL}
\ee
which coincides (again, up to a similar renormalization of couplings)
with the Yang-Lee string equation \rf{gpse} upon $t_3 \sim x_{YL}$, $t_1 \sim t_{YL}$ and
$X_0 = 2\ {\d^2\F\over \d t_1^2} \sim v_{YL}$.

Comparing \rf{koyl} with \rf{kois} one finds that, as we have already done in the previous
section, one may indeed identify $u_I$ with $v_{YL}$ after appropriate shifts of the
variables \rf{renoc} and pointing out the change of the quantum numbers: $t_1 \sim x_I \sim t_{YL}$
and $t_3 \sim t^2_I \sim x_{YL}$.

The relation $u_{YL} \sim v_{YL}^3$ is quite clear from the point of view of
equations \rf{252d}. It is just a particular Hirota equation for dispersionless KP
hierarchy, expressing
\be
u_{YL} \sim\ {\d^2\F\over \d t_3^2} = 3\left({\d^2\F\over \d t_1^2}\right)^3 \sim v_{YL}^3
\ee
the function satisfying equation \rf{Koeq}, and being here a double derivative of
free energy w.r.t the {\em third} time of the hierarchy, in terms of the canonical KP
potential, being always a double derivative w.r.t. the first time.

From the point of view of KP theory it is also rather clear, why equation \rf{Koeq}
is applicable only for $p<q<2p$, in particular only for $K=2,3$ with $\nu(2,2K-1) =
{2\over 2K-3}$. When transforming it to conventional KdV string equation \rf{godeq},
like it was done in \rf{koyl} for the Yang-Lee model, one finds that the variable
$t$ should be generally identified with $t_{7-2K}$-th time of KP hierarchy, which does
not have clear sense at $K>3$.

\subsection{Zamolodchikov curve for Ising}

If all parameters of the gravitational Ising model are ``alive'', the best way
is to study, following \cite{Zamol}, the fixed area partition function
\be
\label{Za}
Z_a = \int{dx\over 2\pi i a} ue^{xa} = - \int{du\over 2\pi i a^2}\ e^{xa}
\\
x = u^3 + {3\over 2} Tu^2 + {H^2\over (u+T)^2}
\ee
where we use again the rescaled variables $x=x_I \sim t_1$, $H \sim t_2$, $T \sim t_5$
and the rescaled Boulatov-Kazakov equation \rf{kbeq} for $u\sim {\d^2\F\over \d x^2}$. The saddle point
equation ${dx\over du}=0$ for the integral in \rf{Za} is given by
\be
\label{spza}
u(u+T)^4 = {2\over 3}H^2 \equiv \xi^2 T^5
\ee
In rescaled variables $u\sim TU$, the saddle point equation \rf{spza} presents a Riemann surface
\be
\label{zc}
U(U+1)^4 = \xi^2
\ee
which is a double-cover of the $U_c$-plane and a 5-sheet cover of the ``magnetic'' plane $\xi$,
and the function $x$ on the curve \rf{zc} contains description of all singularities in the
gravitational Ising model \cite{Zamol_up}.

In particular, the Yang-Lee singularity arises at critical value of magnetic field $H=H_c$,
where two values $\xi_c = \sqrt{2\over 3}{H_c\over T^{5/2}} = \pm i{16\over 5^{5/2}}$
correspond to two points on the curve \rf{zc}
\be
U_c=-{1\over 5},\ \ \ \ \xi_c^2 = U_c(U_c+1)^4 = - {4^4\over 5^5}
\ee
where $\left.{d\xi\over dU}\right|_{U=U_c}$=0. At the Yang-Lee point one also obviously
has $\left.{d^2x\over dU^2}\right|_{U=U_c}=0$, and
\be
\label{xyl}
x = -{7\over 50}T^3 + {5\over 2}(U-U_c)^3 + \ldots = x_c +
{625 \over 128}T^3\left({\xi^2-\xi_c^2\over 2}\right)^{3/2} + \ldots
\ee
so that $X \sim {x-x_c\over T^3} \sim \mu^{3/2}$ scales as the right fractional power of the cosmological
constant $x_{YL}=\mu \sim \xi-\xi_c$ in the Yang-Lee model, corresponding to the well-known scaling
$t_1 \sim t_3^{3/2}$ of KP times at the critical point with $p+q=7$. For the expansion of
Boulatov-Kazakov equation one can now write
\be
\label{xcyl}
{X\over\epsilon^3} \sim \mu V+V^3 + O(\epsilon)
\ee
where $H-H_c\sim\epsilon^2\mu$, $U+{1\over 5}\sim\epsilon V$, i.e. the Zamolodchikov curve in
the vicinity of the Yang-Lee singularity is described, up to renormalization of parameters, by
string equation \rf{gpse}.

\section{Discussion}

We have tried to demonstrate in this paper, that all spherical correlation functions
in quantum Liouville gravity are all contained and can be easily extracted from
the ``science of polynomials'' - dispersionless KP hierarchy. A simple collection of
residue formulas allows to extract the invariant ratios, to be further compared with
the correlation functions in world-sheet theory, which can be now also computed - though
in a much more cumbersome way - mostly due to the results of Alesha Zamolodchikov in
Liouville theory.

Such application of classical integrable science to the problems of two-dimensional
quantum physics is already a step towards dynamical physics from topological strings,
where similar science has been already used with visible success, see e.g.
\cite{WDW,Kontsevich,GKM,EY,LMNMN,tak}.

The most nontrivial point in application of this ``integrable science'' is its
interpretation in terms of the world-sheet theory. The first point concerns resonances
\cite{MSS,BZ}, which allow nonlinear relations between the couplings in KP and world-sheet
theories when the fractions of the KPZ scaling dimensions of couplings \cite{KPZ} are integer.
We have observed this phenomenon in the cases of $(2,7)$ and $(3,4)$ minimal theories, where
it is yet quite easy taken into account by just playing with the residue formulas for one- two-
and three-point functions.
Naively, these resonance reparameterizations look as particular $W$-flows in the space of couplings,
but this question deserves further investigation.

Another nontrivial point is related to the fact that KP times may have different physical sense,
or different quantum numbers, when we expand a KP solution around different backgrounds, corresponding
to particular minimal theories. We have noticed this, say, when comparing the formulations
for the gravitational Ising and Yang-Lee models. This is the simplest observation for a very important
generic fact that physical observables may change their quantum numbers, when effective field
theory is moved in the moduli space - like in four-dimensional supersymmetric gauge theories
electrically charged objects may capture magnetic charges and vice versa. Two dimensional quantum
gravity is therefore a good laboratory for studying such effects.

Finally, let us say few words, how the picture of dispersionless KP for
the minimal string theory could be deformed
towards quasiclassical hierarchies of generic nonsingular type.
An invariant way to look at the basic polynomials $X=\lambda^p+\ldots$ and $Y=\lambda^q+\ldots$
\rf{polspq} is to say, that they satisfy an algebraic equation
\be
\label{pqcurve}
Y^p - X^q - \sum f_{ij}X^iY^j = 0
\ee
with some particular coefficients $\{ f_{ij}\}$. Generally, for arbitrary coefficients
this is a smooth curve of genus
\be
g = {(p-1)(q-1)\over 2}
\ee
which is a resolution or desingularization of the situation, when $X$ and $Y$ can be
parameterized as polynomial of a uniformizing global variable $\lambda$. This number coincides
with the number of primaries in corresponding minimal conformal $(p,q)$ theory.
Such curves
can be obtained, say, by reduction of the curve of two-matrix model \cite{KM}. An
interesting example is the hyperelliptic curve of the $(2,7)$ model $Y^2=X^7+\sum_{j=0}^5f_jX^j$,
satisfied by \rf{pol27} with the coefficients $f_j=f_j(X_0,Y_1,Y_3)$, $j=0,\ldots,5$. At vanishing
times (except for $t_9=2/9$, see \rf{txy27}), it shrinks to a cusp $Y^2=X^7$, or
$Y^2=X^3\left(X^2+{5t_5\over 2}\right)^2$ for nonvanishing cosmological $t_5$,
which is however ``resolved''
by passing to worldsheet times $t_1\to t_1+{25\over 28}t_5^2$ as
$Y^2=(X-u)\left(X^3+{u\over 2}X^2-{u^2\over 2}X-{u^3\over 8}\right)^2$
with $u^2=-{20 t_5\over 7}$. This form directly generalizes the curve of the
Yang-Lee model $Y^2=(X-u)\left(X^2-{u\over 2}X+{u^2\over 4}\right)^2$ for \rf{pol25}
with $u^2=-{12t_3\over 5}$ proportional to nonvanishing cosmological time.

For such curves, the residue formulas we have discussed above should be extended by
period integrals $\oint YdX$ along all nontrivial cycles on the curve \rf{pqcurve}. The
sense of such period integrals is analogous to the Seiberg-Witten periods or the filling fractions
in matrix models. As usually in quasiclassical hierarchies, the appearance of
corresponding period variables reflect increasing number of unfrozen coefficients in the equation
\rf{pqcurve} or new deformations of the background of the minimal string theories. The
study of such deformations is again a long-standing, but still an open problem.

\noindent
{\bf Acknowledgements}. Most of my understanding of the delicate issues
of the relation among different two-dimensional quantum gravity models
has come after conversations with Alesha Zamolodchikov, possessing very deep
feeling of the subject and amazing intuition.
I am also grateful to A.~Belavin, S.~Kharchev, I.~Krichever, A.~Zabrodin and A.~Zamolodchikov
for useful discussions.
The work was partially supported by the Federal Nuclear Energy Agency, the RFBR grant
08-01-00667,
the grant for support of Scientific Schools LSS-1615.2008.2,
the INTAS grant 05-1000008-7865, the project ANR-05-BLAN-0029-01, the
NWO-RFBR program 047.017.2004.015, the Russian-Italian RFBR program 06-01-92059-CE, and by the
Dynasty foundation. I am also grateful for warm hospitality to SUBATECH in Nantes and
IHES in Bur-sur-Yvette, where this work has been completed.


\begin{thebibliography}{7799}


\bibitem{Pol81}
A.Polyakov, Phys.Lett. {\bf B103}(1981) 207-210;
Phys.Lett. {\bf B103}(1981) 211-214.
%
\bibitem{KPZ}
A.Polyakov, Mod.Phys.Lett. {\bf A2} (1987) 893-898;\\
V.Knizhnik, A.Polyakov and A.Zamolodchikov, Mod.Phys.Lett. {\bf A3} (1988) 819-826.
%
\bibitem{DDK}
F.David, Mod.Phys.Lett. A3 (1988) 1651-1659;\\
J.Distler and H.Kawai, Nucl.Phys. B231 (1989) 509-528.
%
\bibitem{mamo}
J.Ambj{\o}rn, B.Durhuus and J.Fr\"olich, Nucl.Phys. {\bf
B257[FS14]} (1985) 433;\\
F.David, Nucl.Phys. {\bf B257[FS14]} (1985) 45;\\
V.Kazakov, Phys.Lett. {\bf 150B} (1985) 282;\\
V.Kazakov, I.Kostov and A.Migdal, Phys.Lett. {\bf 157B} (1985) 295.
%
\bibitem{ds}
V.Kazakov, Mod.Phys.Lett. {\bf A4}(1989) 2125;\\
E.Brezin and V.Kazakov, Phys.Lett. {\bf B236}(1990) 144;\\
M.Douglas and S.Shenker, Nucl.Phys. {\bf B335}(1990) 635; \\
D.Gross and A.Migdal, Phys.Rev.Lett. {\bf 64}(1990) 127.
%
\bibitem{Douglas}
M.Douglas, Phys.Lett. {\bf B238}(1990) 176.
%
\bibitem{FKN}
M.Fukuma, H.Kawai and R.Nakayama, Int.J.Mod.Phys. {\bf A6} (1991) 1385-1406;
Comm.Math.Phys. {\bf 143} (1992) 371-403.
%
\bibitem{MM2}
A.~Marshakov,
  Theor.\ Math.\ Phys.\  {\bf 147} (2006) 777
  [Teor.\ Mat.\ Fiz.\  {\bf 147} (2006) 399]
  [arXiv:hep-th/0601214].
%
\bibitem{GoLi}
M.~Goulian and M.~Li, Phys. Кумю Lett. {\bf 66} (1991) 2051.
%
\bibitem{DOZZ}
H.~Dorn and H.~J.~Otto,
  Nucl.\ Phys.\  B {\bf 429} (1994) 375
  [arXiv:hep-th/9403141];\\
A.~Zamolodchikov and Al.~Zamolodchikov,
  Nucl.\ Phys.\  B {\bf 477} (1996) 577
  [arXiv:hep-th/9506136].
%
\bibitem{Zam_ho}
  Al.~Zamolodchikov,
  Int.\ J.\ Mod.\ Phys.\  A {\bf 19S2} (2004) 510
  [arXiv:hep-th/0312279].
%
\bibitem{BeZa}
A.~Belavin and Al.~Zamolodchikov, in hep-th/0510214.
%
\bibitem{KriW}I.~Krichever,
Commun. Pure. Appl. Math. {\bf 47} (1992) 437, hep-th/9205110.
%
\bibitem{LGGKM}
S.~Kharchev, A.~Marshakov, A.~Mironov and A.~Morozov,
  Mod.~Phys.~Lett.  {\bf A8} (1993) 1047,
  [arXiv:hep-th/9208046].
%
\bibitem{TakTak}
K.~Takasaki and T.~Takebe,
  Rev.\ Math.\ Phys.\  {\bf 7} (1995) 743
  [arXiv:hep-th/9405096].
%
\bibitem{KhMa}
S.~Kharchev and A.~Marshakov,
in
{\sl String Theory, Quantum Gravity and the Unification of the Fundamental
Interactions}, Proceedings of
International Workshop on String Theory, Quantum Gravity and the
Unification of Fundamental
Interactions, Rome, Italy, 1993,  World Scientific;
arXiv:hep-th/9210072;
Int.\ J.\ Mod.\ Phys.\ A {\bf 10}, 1219 (1995)
[arXiv:hep-th/9303100].
%
\bibitem{Kontsevich}
M.~Kontsevich, Func.~Analysis~\&~Apps. {\bf 25} (1991) 50;
Comm. Math.Phys. {\bf 147} (1992) 1
%
\bibitem{GKM}
S.~Kharchev, A.~Marshakov, A.~Mironov, A.~Morozov and A.~Zabrodin,
Phys.\ Lett.\ B {\bf 275} (1992) 311
[arXiv:hep-th/9111037];
Nucl.\ Phys.\ B {\bf 380} (1992) 181
[arXiv:hep-th/9201013].
%
\bibitem{Zamol}
Al.~Zamolodchikov, hep-th/0505063, hep-th/0508044;\\
Al.~Zamolodchikov and Y.~Ishimoto, in hep-th/0510214.
%
\bibitem{BuKa}
V.~Kazakov,
Phys. Lett. {\bf A119} (1986) 140;\\
D.~Boulatov and V.~Kazakov,
Phys. Lett. {\bf B186} (1987) 379.
%
\bibitem{MSS}
G.~Moore, N.~Seiberg and M.~Staudacher, Nucl. Phys. {\bf B362} (1991) 665-709.
%
\bibitem{BZ}
  A.~Belavin and A.~Zamolodchikov,
  arXiv:0811.0450 [hep-th].
%
\bibitem{Zamol_up}
Al.~Zamolodchikov, ``Thermodynamics of Gravitational Ising Model'', 2005, unpublished;\\
 A.~Zamolodchikov and Al.~Zamolodchikov,
  arXiv:hep-th/0608196.
%
\bibitem{KM}
V.~Kazakov and A.~Marshakov,
J.\ Phys.\ A {\bf 36} (2003) 3107
[arXiv:hep-th/0211236].
%
\bibitem{WDW}
E.Witten, Nucl.Phys. {\bf 340} (1990) 281;\\
R.Dijkgraaf and E.Witten, Nucl.Phys. {\bf 342} (1990) 486.
%
\bibitem{EY}
  T.~Eguchi and S.~K.~Yang,
  Mod.\ Phys.\ Lett.\  A {\bf 9} (1994) 2893
  [arXiv: hep-th/9407134];
\\
T.~Eguchi, K.~Hori and S.~K.~Yang,
  Int.\ J.\ Mod.\ Phys.\  A {\bf 10} (1995) 4203
  [arXiv: hep-th/9503017].
%
\bibitem{LMNMN}
  A.~S.~Losev, A.~Marshakov and N.~Nekrasov,
  ``Small instantons, little strings and free fermions,'' in Ian Kogan memorial volume
  {\em From fields to strings:
circumnavigating theoretical physics}, 581-621
  [arXiv: hep-th/0302191];\\
A.~Marshakov and N.~Nekrasov,
  JHEP {\bf 0701} (2007) 104
  [arXiv: hep-th/0612019];\\
A.~Marshakov,
  Theor.\ Math.\ Phys.\  {\bf 154} (2008) 362
  arXiv:0706.2857 [hep-th]; 
  JHEP {\bf 0803} (2008) 055 [arXiv:0712.2802 [hep-th]]; arXiv:0810.1536.
%
\bibitem{tak}
 T.~Nakatsu, Y.~Noma and K.~Takasaki,
  Int.\ J.\ Mod.\ Phys.\  A {\bf 23} (2008) 2332
  [arXiv:0806.3675 [hep-th]]; 
  Nucl.\ Phys.\  B {\bf 808} (2009) 411
  [arXiv:0807.0746 [hep-th]];\\
   T.~Nakatsu and K.~Takasaki, arXiv:0807.4970.








\end{thebibliography}
\end{document}